\documentclass[twocolumn, aps, pra, superscriptaddress]{revtex4-1}

\usepackage{mathptmx}
\usepackage{subfigure}
\usepackage{dcolumn}
\usepackage{amsmath,amssymb}
\usepackage{bm}
\usepackage{color}
\usepackage{latexsym}
\usepackage[english]{babel}
\usepackage{times}

\usepackage{psfrag,graphicx}
\usepackage{epsf}
\usepackage{amsfonts}
\usepackage{natbib}
\usepackage{epstopdf}
\usepackage{appendix}
\usepackage[colorlinks=true,citecolor=blue,linkcolor=blue]{hyperref}

\begin{document}


\title{Deceleration of molecules in a supersonic beam by the optical field in a low-finesse cavity}

\author{Zhihao Lan}
\affiliation{School of Engineering and Physical Sciences, Heriot-Watt University, Edinburgh EH14 4AS, United Kingdom}
\email{lanzhihao7@gmail.com}
\author{Yongkai Zhao}
\affiliation{School of Engineering and Physical Sciences, Heriot-Watt University, Edinburgh EH14 4AS, United Kingdom}

\author{Peter F. Barker}
\affiliation{Department of Physics and Astronomy, University College London, London WC1E 6BT, United Kingdom}
\author{Weiping Lu}
\affiliation{School of Engineering and Physical Sciences, Heriot-Watt University, Edinburgh EH14 4AS, United Kingdom}

\begin{abstract}

We study the dynamics of a supersonic molecular beam in a low-finesse optical cavity and demonstrate that most molecules in the beam can be decelerated to zero central velocity by the intracavity optical field in a process analogous to electrostatic Stark deceleration. We show that the rapid switching of the optical field for slowing the molecules is automatically generated by the cavity-induced dynamics. We further show that $\sim1\%$ of the molecules can be optically trapped at a few millikelvin in the same cavity.

\end{abstract}

\maketitle

The generation of ultracold molecules is opening up new directions in ultracold physics and chemistry. Ultracold molecules offer the possibility of studying exotic quantum phases through anisotropic electric dipole-dipole interactions \cite{r1}. These interactions, however, only become important when they are in the submillikelvin range. The ability to create cold molecules and subsequently trap them in external electric and magnetic fields allows long interaction and interrogation times and therefore high-resolution spectroscopic measurements. Such measurements have already been undertaken with OH radical in the millikelvin temperature range \cite{r2}, which could be used to constrain the time variation in fine structure constant \cite{r3}. Such trapped cold molecules are anticipated to be also important in the search for parity violation \cite{r4}, and for tests of physics beyond the standard model \cite{r5}. Although yet to be explored experimentally, chemistry at ultracold temperatures is dominated by resonance and tunneling phenomena, with reaction rates predicted to be many orders of magnitude larger than at room temperature for some species \cite{r6}.

Cooling of molecules has proven to be considerably more difficult to achieve than laser cooling of atoms \cite{r7}. Ultracold molecules can be created from association of laser-cooled atomic species by photoassociation and on magnetic Feshbach resonances at microKelvin temperatures \cite{r8}. Recently, progress has been made to transfer these molecules in high vibrational levels to low rovibrational states \cite{r9, r10}. The methods are, however, limited to atoms that can be laser cooled. Buffer gas cooling is a general method, which can dissipatively cool complex molecular species. This method utilizes elastic collisions within a buffer gas in a cryogenic cell in the 100-mK range \cite{r11}. Another technique for creating complex cold molecules is via phase space filtering, in which conservative electrostatic, magnetic, or optical potentials are used to filter out a narrow energy distribution of a hotter gas and then transfer them to zero velocity in the laboratory frame \cite{r12, r13, r14, r15, r16}. Electrostatic Stark deceleration is a well-developed scheme of this type where gas of $10^6$cm$^{-3}$ polar molecules in a single quantum state at 10 mK is produced \cite{r12}. This scheme uses rapidly switched electrical fields to create a moving potential that traps and slows a subset of the initial molecular distribution. The rate and duration of the switched field change as the trapped molecules are brought to rest.

In these phase space filtering techniques, the number of slowed molecules reduces with the well depth. For well depths below 1 mK, few molecules can be decelerated and trapped and dissipative cooling is required to increase the phase space density. Techniques such as sympathetic cooling and evaporative cooling offer a route to submillikelvin temperatures. Although less well-developed, cavity cooling also appears feasible to cool atoms or molecules to submillikelvin temperatures. In this scheme an atom strongly coupled to a high Q cavity mode can be cooled below the Doppler limit via a Sisyphus mechanism \cite{r17}. Cooling of many atoms in optical cavities has also been predicted through the correlated dynamics of these atoms \cite{r18}. The investigation has recently focused on the scaling laws of the system with respect to its control parameters in an effort to extend its operation from strong to weak coupling regimes for cooling of a large ensemble \cite{r19, r20, r21}. In this article we study a new optical deceleration scheme in which a supersonic molecular beam is slowed by the conservative optical potential that is automatically switched on and off via the dynamics of the molecule-field interaction in optical cavities. We show that most molecules in the beam are decelerated to zero central velocity. Such a scheme is different from the existing phase space filtering methods but also from cavity cooling.

 \begin{figure}[!hbp]
\centering
\includegraphics[width=1\columnwidth]{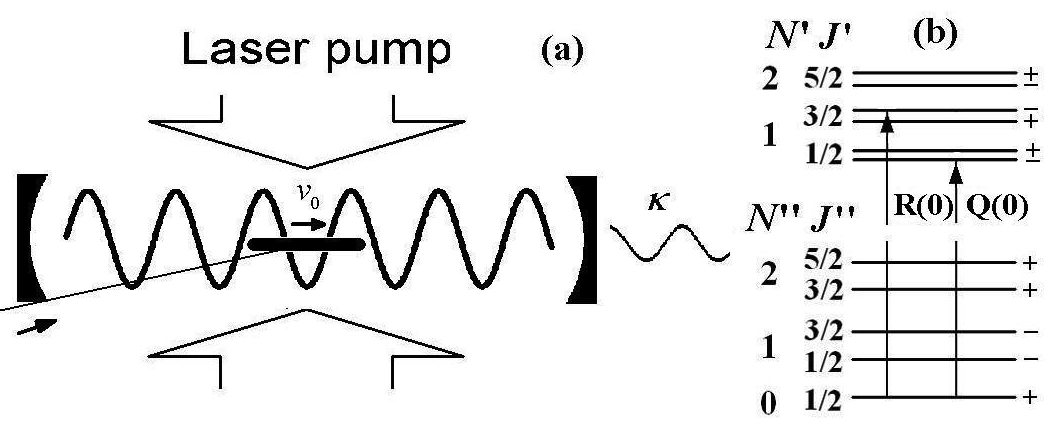}
\caption{ (a) Schematic of an optical cavity-based decelerator. (b) Transition lines of CaF around $606.5$ nm, the $R(0)$ line: $^2\Pi_{1/2}(\nu'=0,N'=1,J'=3/2)\leftarrow X^2\Sigma^+(\nu''=0,N''=0,J''=1/2)$ and $Q(0)$ line: $^2\Pi_{1/2}(\nu'=0,N'=1,J'=1/2)\leftarrow X^2\Sigma^+(\nu''=0,N''=0,J''=1/2)$. }
\label{f1}
\end{figure}

\begin{figure}[!hbp]
\centering
\includegraphics[width=0.8\columnwidth]{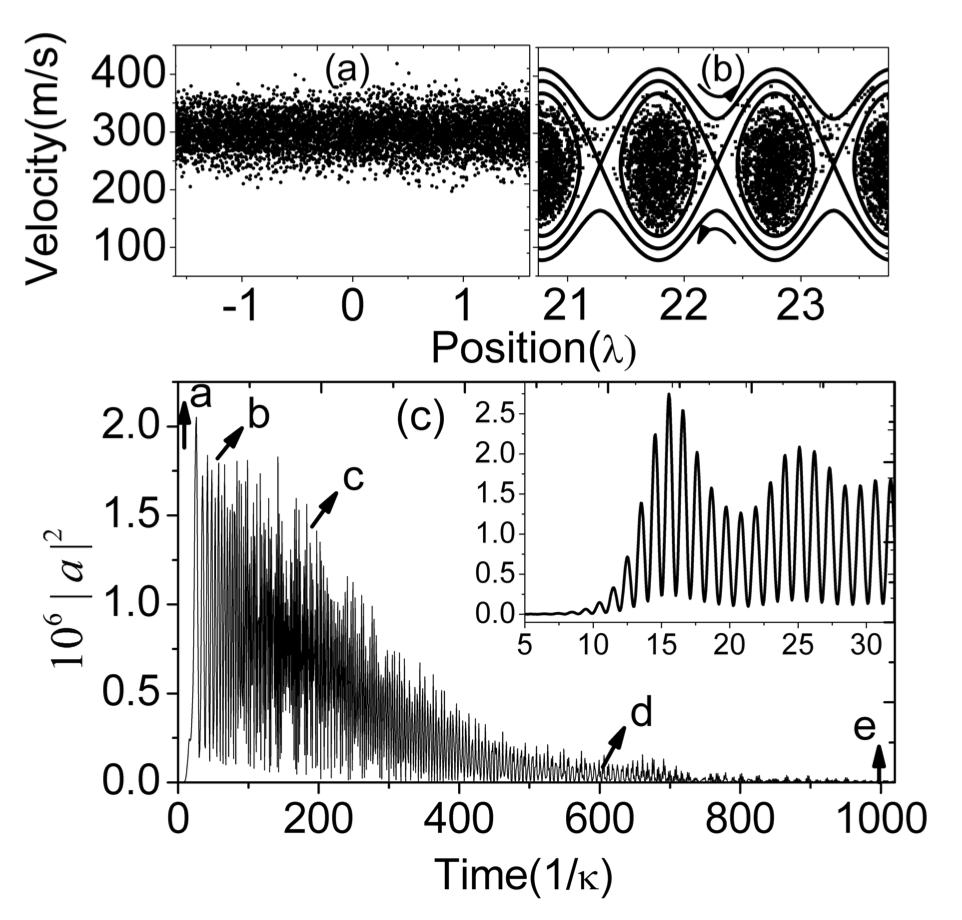}
\caption{  Position-velocity distributions of the molecules at the
initial stage (a) and after spatial self-organization (b). (c) Rapidly
switching optical field intensity in the cavity as the molecular beam
travels along the cavity axis, the insert shows the evolution of the
intensity for the initial period. The parameters used are $g=240$ MHz (the mode value $V=3\times 10^{-6} $mm$^3$), $\Delta_A=-2 \times 10^3$ GHz, $\Delta_c=-10$ GHz, $\Gamma_0=0$, $\kappa=1$ GHz, $\eta=2\times 10^4$ GHz and $N=10^4$.
}
\label{f2}
\end{figure}

We consider a traveling molecular beam entering (nearly axial) an optical cavity that is pumped transversely by laser beams, as depicted in Fig. \ref{f1}(a). The center-of-mass motion of the molecules in the cavity can be described by the well-established semiclassical equations \cite{r18, r19, r20} in one dimension:

\begin{gather}
\dot{\alpha}=(i\Delta_c-\kappa)\alpha-(\Gamma_0+iU_0)\sum_j\cos^2(kx_j)\alpha \nonumber \\ 
-\eta_{\rm{eff}}\sum_j\cos(kx_j)+\xi_\alpha \nonumber \\
\dot{p}_j=\hbar k U_0(|\alpha|^2-1/2)\sin(2kx_j) \nonumber \\
+i\hbar k(\eta^*_{\rm{eff}} \alpha-\eta_{\rm{eff}}\alpha^*)\sin(kx_j)+\xi_{pj}
\label{eq1}
\end{gather}

\noindent
where $\alpha$ is the amplitude of photon number, $p_j$ is the momentum of the $j$th molecule, $ x_j$ is its position so $\dot{x}_j=p_j/m$, $ j=1,\cdots,N$, where $m$ is the mass of the molecules and $N$ is the total number of molecules. The parameters $\Delta_A=\omega-\omega_A$ and $\Delta_c=\omega-\omega_c$ are, respectively, the detuning of the external pump with respect to the molecular and cavity resonances, $U_0=\Delta_Ag^2/(\gamma^2+\Delta_A^2)$ and $\Gamma_0=\gamma g^2/(\gamma^2+\Delta_A^2)$  are dispersion and absorption, where $g$ is the coupling constant and $\gamma$ the molecular spontaneous decay rate, $\eta_{\rm{eff}}=\eta(\Gamma_0+iU_0)/g$ denotes the effective pumping strength for the cavity mode by $\eta$, which is the maximum value of the Rabi frequency of the pump laser (the pump strength). The parameter $\kappa$ is the cavity decay rate and $k$ is the optical wave number. The noise terms, $\xi_{\alpha}$ and $\xi_{pj}$, are of Langevin type, as given in Refs. \cite{r18, r19, r20}. As an example, we consider a cold ($\sim1$ K) supersonic CaF molecular beam. At this temperature, most molecules are in the ro-vibrational ground state and only $R(0)$and $Q(0)$ transitions are allowed for a pump source at around $606.5$ nm [Fig. \ref{f1}(b)]. The system can be approximated by a simple two-level model when the pump field is red detuned from the $Q(0)$ transition and the detuning $\Delta_A$ is much larger than the separation of the two lines ($\sim200$ GHz). The system parameters are given in the Fig. \ref{f2} caption. We choose these parameters, corresponding to a small cavity, in order to compute Eq. (\ref{eq1}) efficiently with $10^4$ molecules. We will later scale up the system to a much larger cavity.

We first apply a similar argument developed in Ref. \cite{r18} for an ensemble of cold atoms to discuss how molecules in the supersonic beam are spatially organized in the optical cavity in the presence of the optical pump field. The initial distribution of the molecules is uniform in space and Gaussian in velocity with central velocity $v_0 = 300$ m/s and half-width $\sigma_0 = 30$ m/s, as shown in Fig. \ref{f2}(a). The traveling molecules being pumped by the external optical field from the transverse direction to the cavity axis scatter photons into the cavity. The amplitude of the scattered field for each molecule depends on the pump field as well as the molecular position. Molecules in the nodes of the standing-wave cavity mode do not make a contribution, whereas those in the antinodes scatter maximally. Since the phase of the coherently scattered field is position dependent, the photons scattered by molecules separated by half a wavelength have opposite phase and interfere destructively, so preventing the buildup of the optical field in the cavity for the uniformly distributed supersonic molecular beam. However, due to random density fluctuations for a finite number of molecules, small optical field in the cavity can emerge momentarily which, for a red-detuned optical pump field, creates an attractive potential to pull molecules to the antinodes of the intracavity field. When the optical pump field exceeds a certain level (threshold), this induced molecular redistribution can strongly enhance the coherent Bragg-type scattering of the pump field into the cavity, which in turn further deepens the optical potential and localizes more molecules in a runaway process. At the end of the process, a traveling molecular density wave is formed with the period equal to the optical wavelength, as shown in Fig. \ref{f2}(b). The formation of such molecular spatial distribution can be explained by the effects of the additional recoil force, compared to the longitudinal pump scheme, on the molecules by the transverse optical pump field. The force is given by the second term of the second equation of Eq. (\ref{eq1}) and has the same period as the spatially organized molecules. Fig. \ref{f2}(b) also shows the potential lines produced by the optical field in the cavity. Since the optical field depends on the position of the molecules, it oscillates as the molecular beam travels along the cavity axis as shown in Fig. \ref{f2}(c), where the insert gives the initial buildup of the optical field in the cavity, which occurs simultaneously with the spatial self-organization process of the molecules.

Molecules in the supersonic beam must be stably bunched in the optical dipole potential wells and experience a dissipative optical force in order to be decelerated along the cavity. The stability of the bunched molecules, as shown in Fig. \ref{f2}(b), can be understood from a simplified model of Eq. (\ref{eq1}). The intracavity optical field strength in the adiabatic limit is from Eq. (\ref{eq1}),
\begin{gather}
|\alpha|^2=\frac{|\eta_{\rm{eff}}|^2\left[\sum_j\cos(kx_j)\right]^2}{\left[ \kappa+\Gamma_0\sum_j\cos^2(kx_j)\right]^2+ \left[ \Delta_c-U_0\sum_j\cos^2(kx_j)\right]^2} \nonumber \\
\approx \frac{|\eta_{\rm{eff}}|^2\left[\sum_j\cos(kx_j)\right]^2}{\kappa^2+\Delta_c^2}
\end{gather}
where we have neglected the term with $U_0$ under the condition $U_0N\ll\Delta_c$ for the latter expression and $\Gamma_0=0$. To
evaluate $\sum_j\cos(kx_j)$, we introduce a distribution function
$f(x-vt)$ to describe the bunched molecules, where $v$ is
the central velocity of molecules in the supersonic beam.
The distribution function is a traveling wave and has a
spatial period of $2\pi$, based on the observation of Figs. \ref{f2}(b) and \ref{f2}(c). So $\sum_j\cos(kx_j)\approx N\int\cos(kx)f(x-vt)dx=N\cos(kvt)\int\cos(kx)f(x)-N\sin(kvt)\int\sin(kx)f(x)dx$. We can choose the initial condition so that the peaks of
$f(x)$ are positioned at $x = 0, 2\pi,4\pi,\cdots $. This leads to
$\int\sin(kx)f(x)dx=0$, so $\sum_j\cos(kx_j)\approx N_{\rm{eff}}\cos(kvt)$ where $N_{\rm{eff}}=N\int\cos(kx)f(x)dx$ is the
effective number, which
depends on the total number of the molecules in the supersonic
beam and their distribution. The optical field intensity in the
cavity is then given as $|\alpha|^2|\approx\eta_{\rm{eff}}|^2N_{\rm{eff}}^2\cos^2(kvt)/(\kappa^2+\Delta_c^2)$
which indeed switches on and off, as shown in Fig. \ref{f2}(c). Under the condition $U_0N\ll \kappa$, the first term on the right-hand side of the second equation of Eq. (\ref{eq1}) can be neglected compared to the second term and the simplified equation of motion is given by
\begin{gather}
\dot{p}\approx=i\hbar k(\eta^*_{\rm{eff}} \alpha-\eta_{\rm{eff}}\alpha^*)\sin(kx)=f\cos(kvt)\sin(kx) \nonumber \\
=(f/2)\left[\sin(kx+kvt)+\sin(kx-kvt)\right]
\label{eq3}
\end{gather}
where $f=2|\eta_{\rm{eff}}|^2\hbar k\Delta_cN_{\rm{eff}}/(\Delta_c^2+\kappa^2)$ is the amplitude of the dipole force. The dynamics of the system is therefore governed by the dipole forces from the two counter-propagating optical waves of velocity $v$ in the cavity. In fact, Eq. (\ref{eq3}) has the same form as that obtained in Stark decelerator when only the first harmonic of the electric field is considered \cite{r22}. As discussed in Stark deceleration, molecules that move at velocities close to that of the wave component in the $+x$ direction interact more significantly with it, and the wave component in the $-x$ direction can be neglected. As such, the important dynamics is reduced to traveling molecules in an optical lattice of same velocity [Fig. \ref{f2}(b)], equivalent to the transportation scheme in Stark deceleration \cite{r22}. The phase stability in our scheme thus results from the cavity-mediated dynamics of the optical field.

\begin{figure}[!hbp]
\centering
\includegraphics[width=1.0\columnwidth]{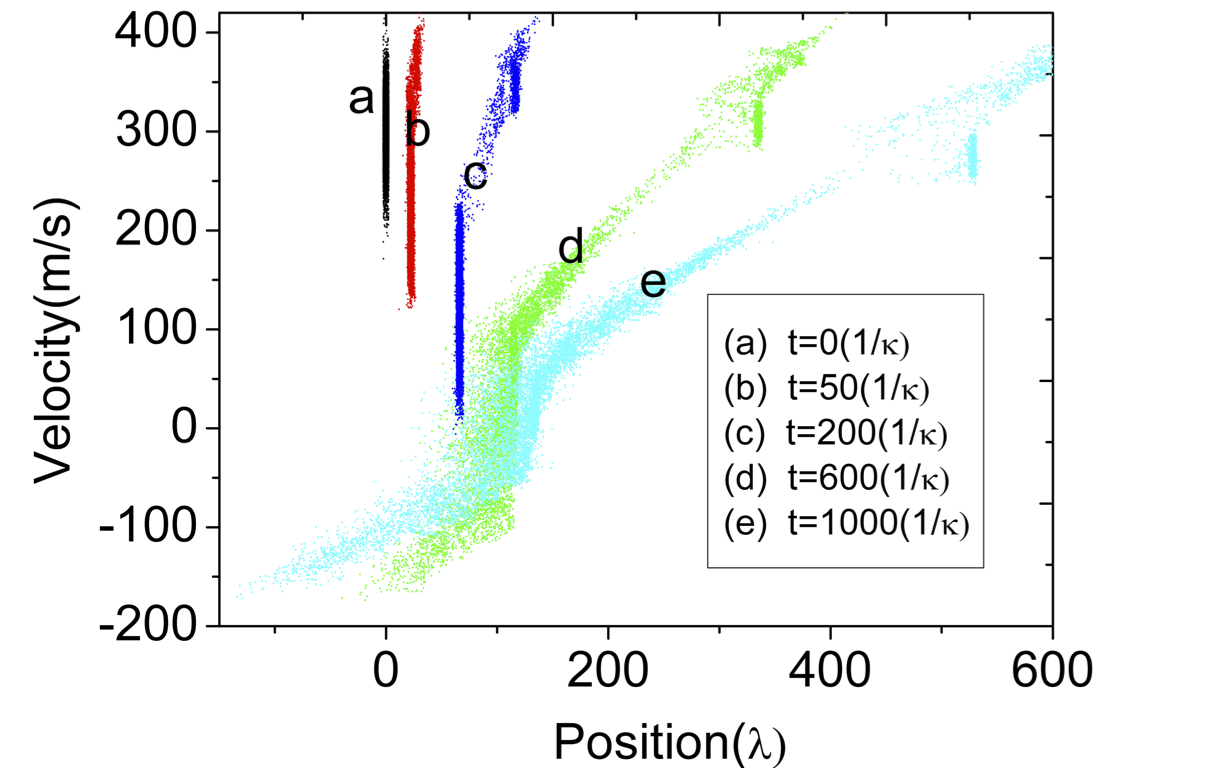}
\caption{ (Color online)  Phase space plot of a traveling molecular beam at different times, traces (a)Ð(e), the intracavity field intensities of which are marked in Fig. \ref{f2}(c). The parameters are the same as in Fig. \ref{f2}.}
\label{f3}
\end{figure}

\begin{figure}[!hbp]
\centering
\includegraphics[width=1.0\columnwidth]{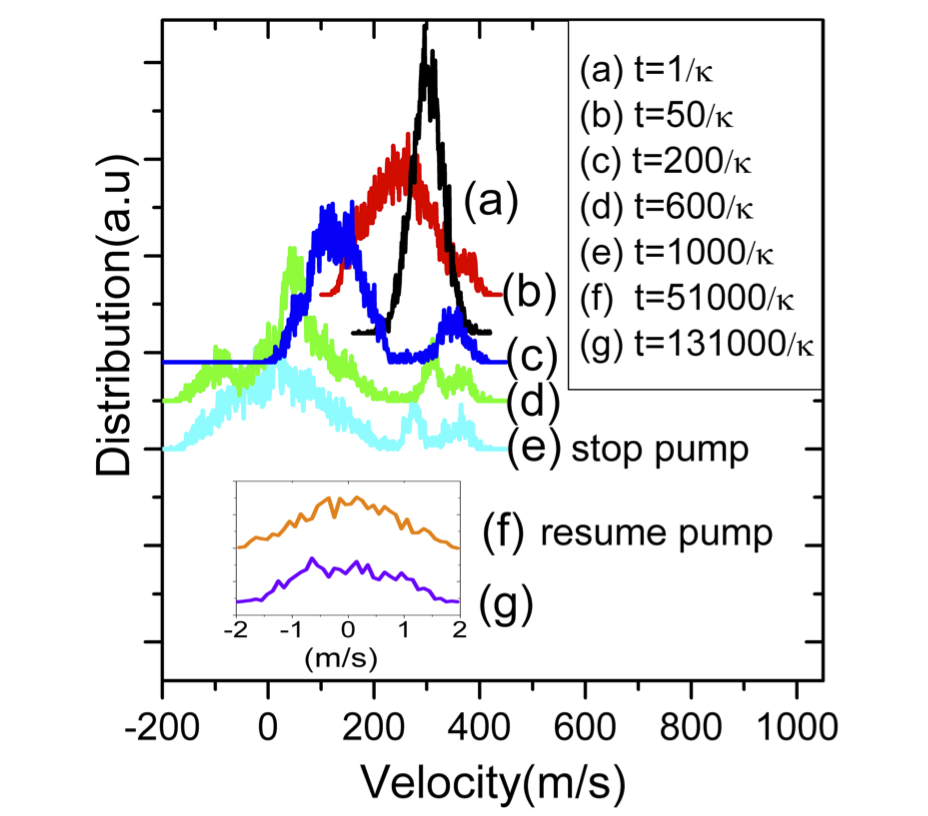}
\caption{ (Color online)  Velocity distributions of the molecules corresponding to trace (a)Ð(e) in Fig. \ref{f3}. Traces (f) and (g) show the initial and final distributions of trapped molecules in the cavity. The parameters used are the same as in Fig. \ref{f2}.}
\label{f4}
\end{figure}

To slow down the molecular beam, a dissipative optical
force is required to act on the molecules. Such a force can
be produced by the nonadiabatic effects of cavity dynamics as
studied extensively in the early work of cavity cooling \cite{r17, r18, r19, r20}.
Therefore, a compromise is needed in the choice of the cavity
parameters if we require both phase stability and molecular
deceleration (dissipation). We have extensively studied the
operation conditions of the system by analyzing the numerical
results of Eq. (\ref{eq1}) and found that the requirement can be met 
by appropriately setting the ratio $r=kv_0/\sqrt{(\kappa^2+\Delta_c^2)}$, where $1/kv_0$ is the time for molecules to travel one wavelength and
$1/\sqrt{\kappa^2+\Delta_c^2}$ the detuning-enhanced cavity lifetime. A smaller
(larger) $r$ means faster (slower) cavity response to the dynamics of molecules, which leads to better (poorer) molecular spatial organization but weaker (stronger) deceleration. We find that deceleration works within the window $0.1 < r < 0.6$. For a supersonic beam of $v_0 = 300$ m/s, we have chosen $\kappa = 1$ GHz and $\Delta_c=-10$ GHz so $r =0.3$ .We note that the value of $\Delta_c$ we have chosen is very different from that for observing cavity cooling of atoms \cite{r18, r19} and no effective cooling occurs for the parameters we set here.

Figure \ref{f3} is the simulation of deceleration of the molecular beam for a pump level of $\eta = 2\times10^4$ GHz, which is some $30\%$ above the threshold. Spontaneous emission from the excited states is weak (the saturation parameter as defined in Ref. \cite{r23}$ \sim1\%$) at this pump level and can be neglected. The traces (a)Ð(e) show the evolution of the molecules in the space-velocity space. As observed, the phase stability is well maintained when the molecules are decelerated until they touch zero velocity [traces (aÐc)], evident by the vertical shape of the bunched molecules in the beam and the nearly constant amplitude of the intracavity field during the period as shown in Fig. \ref{f2}(2). We note that to avoid spatial overlapping in the display of the molecular beam, Fig. \ref{f3} is the simulation for a short molecular beam of only five wavelengths. The results for a long beam of hundreds of wavelengths remain essentially the same, as the boundary effects at the two ends of the beam play little role. Further slowing-down of the molecules from Fig. \ref{f3}, trace (c), leads to the reduction of molecular number in the beam [traces (dÐe)], which in turn decreases the intracavity field intensity, as shown in Fig. \ref{f2}(c). The molecules gradually lose phase stability during the period. This process continues until the intensity drops to zero, where the external optical pump field is switched off. The velocity distributions at different stages are given in Fig. \ref{f4}(a)-\ref{f4}(e). The final distribution has three peaks. The main peak consists of the molecules that have been synchronously slowed to a zero central velocity, the velocity half-width of which is $\sigma \approx 70$ m/s, approximately twice the initial value $\sigma_0$. We note that the half width of the slowed molecules depends on the pump intensity and increases (decreases) with the increase (decrease) of the pump. The two smaller peaks are for nonsynchronous molecules close to velocity $v_0$. Our simulation shows the decelerated molecules travel a distance of around $100 \lambda$ and for duration of $10^3  \kappa^{-1}$ before the central velocity is reduced to zero. In general, this decelerator requires much shorter deceleration time and traveling distance compared to the electrostatic Stark decelerator.

\begin{figure}[!hbp]
\centering
\includegraphics[width=1\columnwidth]{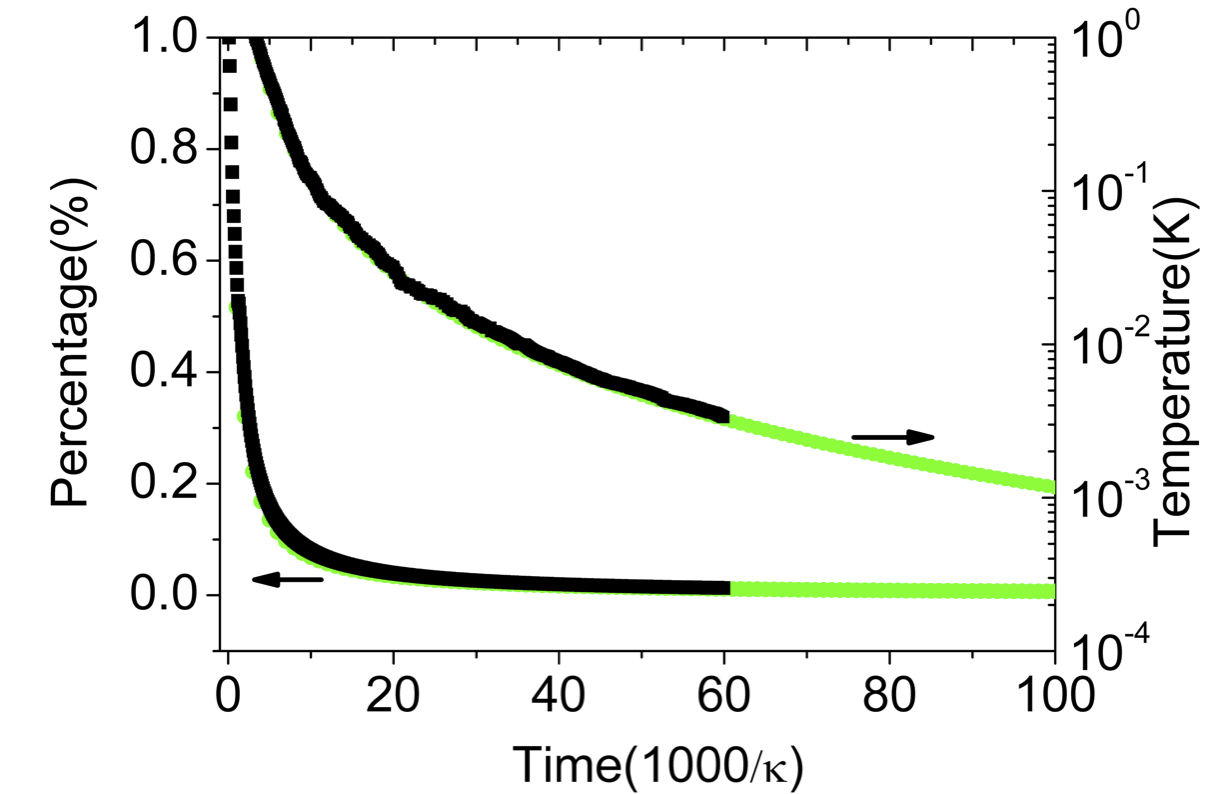}
\caption{ (Color online) The theoretical (green) and simulation (black) results of the percentage of the molecules and their temperatures in the cavity. }
\label{f5}
\end{figure}

\begin{figure}[!hbp]
\centering
\includegraphics[width=1\columnwidth]{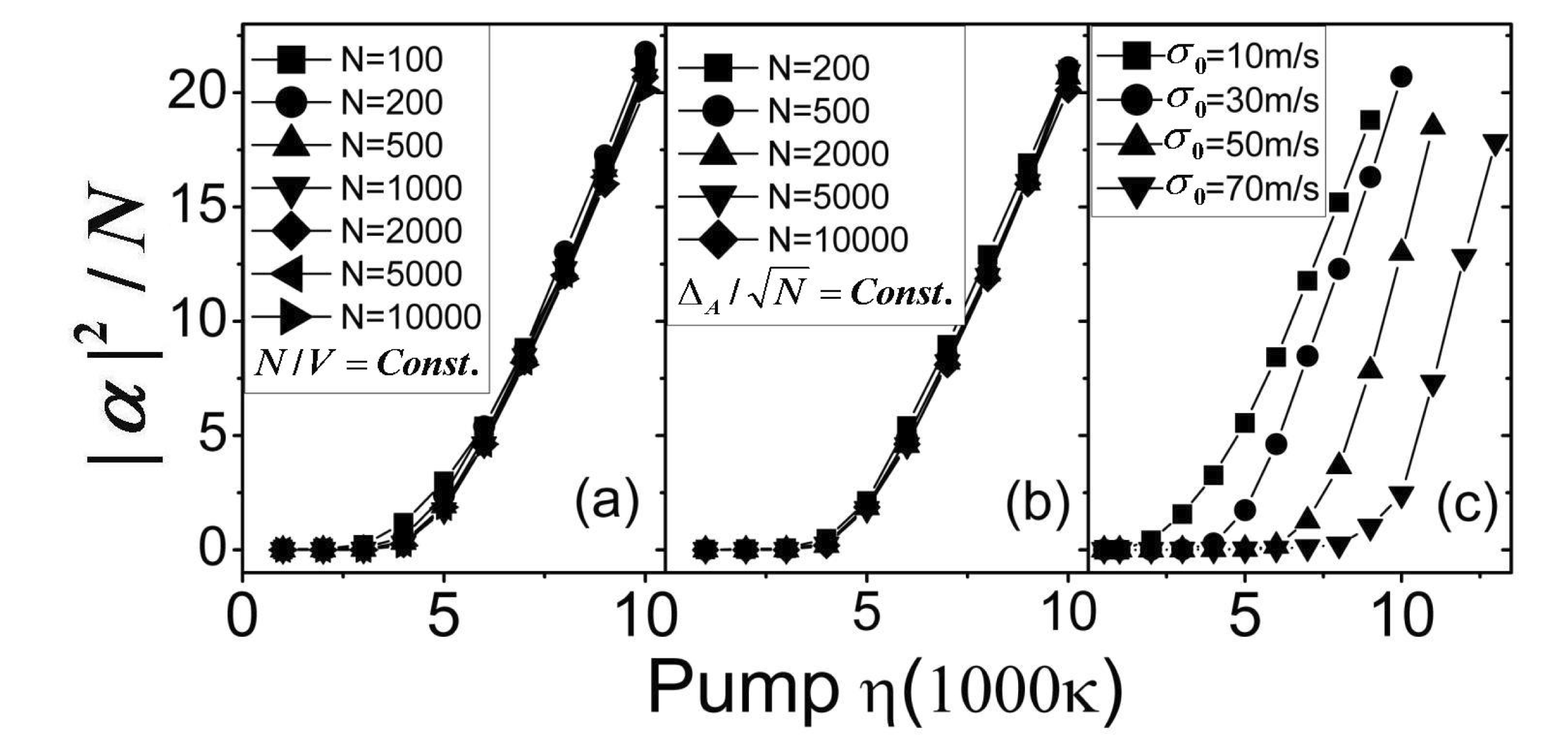}
\caption{ Scaling with respect to (a) $N$ and $V$ (or $1/g^2$) ($Ng^2=5.76\times10^8$ MHz$^2$ fixed), where $\Delta_A=-2\times10^3$ GHz and 
$\sigma_0=30$ m/s, (b) $\Delta_A$ and $\sqrt{N}$ ($\Delta_A/\sqrt{N}=-200$ GHz fixed), where $g=240$ MHz and $\sigma_0=30$ m/s, and (c) $\sigma_0$, where $\Delta_A=-2\times10^3$ GHz and $N=10^4$. Other parameters are $\kappa=10$ GHz, $\Delta_c=-100$ GHz and $v_0=300$ m/s.
 }
\label{f6}
\end{figure}
 
 We now discuss a procedure for optically trapping the decelerated molecules in the same cavity. We assume that the molecular beam enters the cavity at a small angle of around $\theta=5^0$ to the axis and has a beam length of $l=2w_0/\theta$,where $w_0$ is the cavity waist. When the central velocity of the slowed molecules goes to zero, they maintain a Gaussian velocity distribution with $\sigma \approx 70$ m/s [Fig. \ref{f4}(e)] and are largely uniform within ($-l/2, +l/2$) because of very short deceleration time. These molecules move freely along the beam directions in the absence of the optical field in the cavity. After a certain time, only a small percentage of the molecules at the low end of the velocity distribution remain in the cavity. The molecular number can be calculated by integral over the space between $x_1 =-l/2$ and $x_2 =+ l/2$ and the velocity between $v_1 =(x-l/2)/t$ and $v_2 =(x+l/2)/t$,
 \begin{gather}
 N(t)=N\frac{1}{l}\frac{1}{\sqrt{2\pi}\sigma}\int_{x_1}^{x_2}dx\int_{v_1}^{v_2} e^{-\frac{v^2}{2\sigma^2}}dv \nonumber \\
 \approx \frac{N}{\sqrt{2\pi}}\left(\frac{l}{\sigma t}-\frac{l^3}{12\sigma^3t^3} \right),
 \label{eq4} 
 \end{gather}
 where $\exp(-v^2/2\sigma^2)\approx1-v^2/(2\sigma^2)$ has been used. The temperature of the molecules is given by

  \begin{gather}
 T(t)=\frac{N}{N(t)}\frac{1}{l}\frac{m}{\sqrt{2\pi}\sigma k_B}\int_{x_1}^{x_2}dx\int_{v_1}^{v_2}e^{-\frac{v^2}{2\sigma^2}}dv \nonumber \\
 \approx \frac{N}{N(t)}\frac{m}{k_B\sqrt{2\pi}}\left(\frac{l^3}{6\sigma t^3}-\frac{l^5}{30\sigma^3t^5}\right).
 \label{eq5}
 \end{gather}
 The results from Eqs. (\ref{eq4}) and (\ref{eq5}) have been compared directly to the simulation results. These agree well as shown in Fig. \ref{f5}, so the above two expressions can be used to predict the numbers of molecules and their temperatures in the cavity. Figure \ref{f4}, trace (f), shows the velocity distribution of the $\sim 1\%$ of the molecules remains in the cavity at $t = 51\times 10^3 \kappa^{-1}$, the corresponding temperature (energy spreading) being $4.7$ mK. If we turn on the external optical pump field used for slowing the molecular beam at this moment but at a significantly reduced level	of $\eta=300\kappa$, over $80\%$ of the	remaining molecules can be trapped in the cavity, the temperature of which is $4$ mK [Fig. \ref{f4}(g)]. We note that because the cavity operates in the weak coupling region, it acts to trap rather than cool molecules at this stage.

 Next we investigate the scaling of the threshold pump for the onset of intracavity dynamics with respect to the molecular number, cavity size, and pump power. The scaling laws are explored here in order to decelerate and trap a larger molecular sample using our results for the case of a small cavity. We consider the threshold condition as a linear stability problem of the coupled intracavity field and Boltzmann equations \cite{r20}. In doing so, we linearize the coupled equations in the adiabatic limit ($\kappa\rightarrow\infty$) around the steady-state solution (trivial solution: $\alpha=0,  f(x,t)=\exp\left[-(v-v_0)^2/2\sigma_0^2\right]$) and then solve the linearized equations as an eigenvalue problem. When the real part of one of the eigenvalues becomes positive, the steady-state solution is unstable and leads to exponential growth of a traveling wave, the so-called runaway process as mentioned previously. The occurrence of the positive real part of an eigenvalue depends on the system parameters and can be used to determine the threshold pump,
 \begin{gather}
 \eta_{\rm{th}}=\sqrt{\frac{m}{\hbar}}\frac{\Delta_A\sigma_0}{\sqrt{N}g}\sqrt{\frac{\Delta_c^2+\kappa^2}{|\Delta_c|}}.
 \label{eq6}
 \end{gather}
 The full detail will be published separately. Equation (\ref{eq6}) is
consistent with the mean-field approximation \cite{r19} and our
previous work \cite{r20} under the relation $m\sigma_0^2/2=k_BT/2$. Since $g\propto1/\sqrt{V}$ where $V$ is the cavity mode volume, Eq. (\ref{eq6})
indicates that the threshold is unchanged for a fixed ratio $N / V$. We verify Eq. (\ref{eq6}) by plotting the simulation results of $\eta$ vs.
$|\alpha|^2/N$ for fixed $N/V$ and $\Delta_A/\sqrt{N}$ in Figs. \ref{f6}(a) and \ref{f6}(b). Overlapping of the curves establishes that both threshold pump
and system dynamics remains unchanged. Such invariance can indeed be obtained from Eq. (\ref{eq1}) under the conditions of
$\sum_j\cos(kx_j)\propto N$,which is valid when molecules are spatially organized. Figure \ref{f6}(c) shows $\eta_{\rm{th}}\propto \sigma_0$. The near parallelism of the curves in this figure indicates a good linear scaling for different $\sigma_0$ as long as $\sigma_0$  is much smaller than $v_0$. In relating the above results to practical operation of the system, we further find that these scaling laws remain valid once the ratio r is set close to the lower end of the window $0.1 < r < 0.6$, where the adiabatic effect is largely maintained.
 
 Finally, we use Eq. (\ref{eq6}) to scale up the small cavity. For simplicity, we increase both $V$ and $N$ by a factor of $1.28 \times 10^7$ while keeping other parameters the same. Figures \ref{f2}, \ref{f3}, and \ref{f4} now correspond to the results of $1.28 \times 10^{11}$ CaF molecules (density of $8 \times 10^{12}$ cm$^{-3}$ \cite{r12}) in a supersonic beam of $2$ cm in length and $1$ mm in diameter. The optical cavity now has the length of $5$ cm, the waist of $1$ mm, the finesse of $9.4$, and $g = 0.067$ MHz. The operation pump power is $P =2\kappa \eta^2\hbar\omega=0.26$ W for the duration of $1\mu$s. The time required for escape from the cavity of hotter molecules in the decelerated beam is $11.7$ ms, based on the scaling in Eqs. (\ref{eq4}) and (\ref{eq5}). A pump power required to optically trap the slowed molecules is $60 \mu$ W. The density and temperature of the trapped molecules are $6.8 \times 10^{10}$ cm$^{-3}$ and $4$ mK. We note that when the cavity length is increased to $5$ cm, its free spectral range (FSR) is reduced to $3$ GHz, which is shorter than the frequency detuning $\Delta_c=-10$ GHz. To maintain single-mode operation, a Fabry-Perot etalon should be inserted to the cavity, which is a standard method used in optical devices.
 
In conclusion, we have shown that the interplay of the optical pump with a supersonic molecular beam in an optical cavity can produce two dynamical effects: it segments the beam into a periodic density wave and generates an intracavity optical field via coherent Bragg-type scattering. The optical field is then switched dynamically with the traveling molecules in each cycle of the cavity mode. The nonadiabatic nature of the cavity dynamics gives rise to a friction force, which slows most molecules to zero central velocity through many cycles. We note that our scheme is fundamentally different from that of microwave Stark decelerator \cite{r24}, in which a time-varying external field is required and only a small percentage of molecules can be slowed because of the absence of molecule-cavity interaction. Our scheme works in a low-finesse cavity with large cavity detuning, which is also very different from that for cavity cooling of cold atoms and microwave Stark decelerator. Finally, we find that our scheme can operate in a broad parameter window, is scalable, and is realistic for experiments.


\end{document}